\documentclass[epj]{svjour}
\usepackage[T1]{fontenc}
\usepackage[latin9]{inputenc}
\usepackage{verbatim}
\usepackage{amstext}
\usepackage{color}
\usepackage{graphicx,psfrag}
 \usepackage{amssymb}

\def\beq{\begin{equation}}
\def\eeq{\end{equation}}
\def\beqa{\begin{eqnarray}} 
\def\eeqa{\end{eqnarray}}
\def\beg{\begin{lyxgreyedout}}
\def\eeg{\end{lyxgreyedout}}

\begin{document}

\title{On the Compton clock and the undulatory nature of particle mass in graphene systems}
\author{Donatello Dolce\inst{1} \and Andrea Perali\inst{1}}
\institute{University of Camerino, Piazza Cavour 19F, 62032 Camerino, Italy.}


\abstract{In undulatory mechanics the rest mass of a particle is associated to a rest periodicity known as Compton periodicity. In carbon nanotubes the Compton periodicity is determined geometrically, through dimensional reduction, by the circumference of the curled-up dimension, or by similar spatial constraints to the charge carrier wave function in other condensed matter systems. In this way the Compton periodicity is effectively reduced by several order of magnitudes with respect to that of the electron, allowing for the possibility to experimentally test foundational aspects of quantum mechanics. We present a novel powerful formalism to derive the electronic properties of carbon nanotubes, in agreement with the results known in the literature, from simple geometric and relativistic considerations about the Compton periodicity as well as a dictionary of analogies between particle and graphene physics.
\PACS{
      {61.46.Fg}{Nanotubes}   \and
      {03.65.Sq}{Semiclassical theories and applications} 
     } 
} 

\authorrunning{D. Dolce \& A. Perali}
 \titlerunning{On the Compton clock and the undulatory nature of particle mass in graphene systems}
\maketitle

\section*{{Introduction}}

The origin of the fundamental properties of elementary particles  represents one of the most challenging aspects of modern physics. For instance the investigation of the origin of particle masses is the main task of modern particle collider experiments such as LHC. Nevertheless some of these fundamental aspects  can be indirectly explored in a very accessible way through graphene physics.
Typical examples are the emergence of the Dirac equation and the (pseudo)spin degree of freedom from the geometry of the graphene sub-lattice or of the (pseudo)magnetic field from lattice deformations \cite{RevModPhys.81.109,PhysRevLett.106.116803}. 
  Indeed graphene allows us to simulate and manipulate (low dimensional) space-time geometries, and therefore to reproduce in test-tubes fundamental geometric aspects of some of the most advanced high energy physics theories. 
 One of the aims of this paper is to establish, in terms of geometric considerations,  a dictionary of  correspondences between fundamental aspects of graphene and particle physics, summarised in Tab.(\ref{tab1}), with particular emphasis on particles mass.  In this way we provide a further phenomenological test in graphene systems of the equivalence between the quantisation method proposed in \cite{Dolce:cycles,Dolce:tune,Dolce:AdSCFT,Dolce:2009ce}, see also \cite{tHooft,hooft:2014cogwheel} and ordinary quantum mechanics.

 The low energy excitations of Elementary Charge Carriers (ECCs) in a graphene layer behave as massless (quasi) particles, in which the speed of light $c$ is replaced by the Fermi velocity $v_F$. Nevertheless, in graphene systems such as Carbon Nanotubes (CNs) and bilayer graphene (BL), a finite effective mass is generated for the (pseudo)particles. 
The origin of this effective mass can be easily interpreted in terms of the rest quantum recurrence prescribed by the wave-particle dualism for massive particles. As discovered by de Broglie \cite{Broglie:1924}, elementary particles are ``periodic phenomena'' whose recurrences determine their kinematics through the Planck constant. Thus in undulatory mechanics the mass $\bar m$ (rest energy) is related to a rest recurrence in time, namely the Compton periodicity $T_C = h / \bar m c^2$, also known as the de Broglie internal clock or Compton clock. For Dirac particles this corresponds to the \emph{zitterbewegung} (i.e. ``trembling motion'' \cite{PhysRevB.80.045416}). Indeed, as also pointed out by Penrose \cite{Penrose:cycles}, the rest recurrence of massive elementary particles suggests that  \emph{every isolated particle is a relativistic reference clock}  \cite{Dolce:cycles,Dolce:tune,Dolce:AdSCFT,Dolce:2009ce}, i.e. ``a clock directly linking time to a particle's mass'' \cite{Lan01022013}.  

In graphene systems the Compton periodicity is directly determined by the geometry of the graphene system. In particular the circumference $C_h$ of the curled up dimension of a CN explicitly represents the Compton wavelength of the system. In fact, a quasi-particle at rest with respect to the axial direction of the CN, moves at velocity $v_F$ along the circumference.  Thus, through dimensional reduction of a graphene layer in order to form a CN, the  the quasi-particle acquires a finite rest recurrence, i.e. a Compton periodicity $T_C = \frac{C_h}{v_F}$ with respect to the one dimensional effective motion along the axial direction. Hence the mass scale of the ECCs in a CN is given by the Compton relation $\bar m = {h}/{T_c v_F^2}$. Contrarily to the Compton periodicity of the electron which is about $10^{-21} s$,  in CNs the effective Compton periodicity is rescaled to about $10^{-15} s$, so that it can be directly accessible by modern experimental techniques (e.g. interferometric techniques) \cite{Lan01022013,Margolis_2014,Rusin:ZBG:CNs,Catillon:DebroglieClock}.  Thus CNs can by used to experimentally test foundational aspects of quantum mechanics (QM) related to the Compton periodicity, as pointed out in recent publications \cite{Dolce:cycles,Dolce:tune,Dolce:AdSCFT}.  This also suggests for the possibility to use CNs as reference clocks of precision  $\sim 10^{-15} s$ \cite{Lan01022013}. The formalism developed here to study the quantum recurrences  of ECCs could also be useful to manipulate information in CNs as ultra-fast electronic devices.  

In this paper we will present  an heuristic derivation of electronic properties, such as the energy bands, of CNs from simple considerations about the relativistic modulations of the ECCs Compton clock, \cite{Dolce:cycles,Dolce:tune,Dolce:AdSCFT,Dolce:2009ce} $T_C = {C_h}/{v_F}$.  The results obtained are in full agreement with the ordinary CN description based on standard quantum mechanics, Bloch's theorem and hopping terms. Although the results presented here are not novel, our description can be seen as an intuitive method for a quick qualitative identification of the essential electronic properties of CNs. This can be particularly useful to investigate more complex graphene geometries or other condensed matter systems, such as graphene bi-layers or superlattice of quantum stripes, also described in the text. 

Summarising the idea, as a consequence of the quasi-particle motion with generic momentum $\bar p_{\parallel}$ along the CN, the periodicity of the Compton clock is modulated in analogy with the relativistic Doppler effect. Indeed, the Lorentz transformation and the de Broglie - Planck - Einstein relation of undulatory mechanics  implies that the resulting modulated time recurrence is $T(\bar p_{\parallel}) = h (\bar m^2 v^4_F + \bar p_{\parallel}^2 v_F^2)^{-1/2}$, see \cite{Dolce:cycles,Dolce:tune,Dolce:AdSCFT,Dolce:2009ce}, such that $T(0) = T_C$. The condition of time periodicity yields a quantized energy spectrum $E_n (\bar p_{\parallel}) = n h / T(\bar p_{\parallel}) = n h (\bar m^2 v^4_F + \bar p_{\parallel}^2 v_F^2)^{1/2}$, similarly to the harmonic frequency spectrum $f_n = n / T$ of a string vibrating with fundamental frequency $f = 1 / T$. This represents the energy bands of a CN in the limit of continuum lattice. Its generalisation to relativistic space-time corresponds to the normal ordered energy spectrum prescribed by the second quantisation for a free relativistic bosonic particle. The effective bands of a CN can be then derived by considering the periodicity on a lattice. As a homogenous chain of $N$ springs and masses  has frequency spectrum $f_n = \frac{N}{T \pi} \sin\frac{n \pi}{ N}$, way we find that, for instance, the mass spectrum (rest energy) of a Armchair CN is $m_n = \frac{h}{T_C v_F^2 }\frac{N}{\pi} \sin \frac{n \pi}{N}$, in agreement with \cite{RevModPhys.79.677,deWoul:2012ed}. Similarly, according to Compton relation, the mass of the fundamental band of a metallic CN is $m^* = h /{T_C v_F^2}$, whereas for a semiconducting CN is $m^* = h / {3 T_C v_F^2}$ due to a twist factor of $2 \pi /3$ in the condition of periodicity resulting from the hexagonal lattice geometry. In short, the energy bands can be regarded as the modulated vibrational spectra of Compton clocks on the honeycomb lattice.

 In \cite{Dolce:superc} we have used a similar consideration about the quantum recurrences to derive the fundamental phenomenology of  superconductivity and gap opening with particular reference to CNs, obtaining an heuristic interpretation of the prediction that the critical temperature is inversely proportional to the CN circumference \cite{WhiteCNs}.

\section{{Effective mass and Compton periodicity}}

The quantum recurrence of the ECCs wave function in a graphene layer is determined by the periodicities of the honeycomb lattice. By assuming the base vectors $\vec a_{1,2} = (\sqrt 3, \pm 1) a / 2$, where $a \simeq 2.46 $ \AA ~  is the lattice spacing, the periodicity of the effective graphene space-time, characterized by two spatial dimensions (2D), is $\vec r \equiv \vec r + \vec a_1 n + \vec a_2 m$, with $n, m$ $\in \mathbb N$.  Thus the ECCs wave function fulfills the following Periodic Boundary Conditions (PBCs):
  \beq\psi(\vec r) \equiv \psi(\vec r + \vec a_1 n + \vec a_2 m)\,.\label{graphene:lattice}\eeq 
  According to Bloch's theorem, these PBCs determine the behaviors of the ECCs in the  graphene layer.  In this paper, for the sake of simplicity, we will not consider the pseudo-spin degrees of freedom of the ECCs, whose geometric interpretation in terms of graphene sub-lattice is well-known in literature \cite{RevModPhys.79.677}. 
  
 In undulatory mechanics the energy $\bar E(\bar \mathbf p)$ and momentum $\bar \mathbf p$ of a particle are determined by the time and spatial recurrences through  the de Broglie - Planck - Einstein relations: 
  \beq
  \bar E(\bar \mathbf p) = \frac{2 \pi \hbar}{ T (\bar \mathbf p)}~, ~~~~~~~~~~~~ \bar p_i = \frac{2 \pi \hbar}{  \lambda_i }~. \label{deBroglie:Planck:rel}
\eeq 
with $i=1, 2$ in the case of the effective (2D+1) space-time associated to the graphene layer. Roughly speaking, in the wave-particle duality the energy-momentum and the space-time recurrences are ``two faces of the same coin''.

Graphene systems suggest an undulatory description of the (quasi)particle effective masses. As a consequence of the honeycomb lattice periodicity (\ref{graphene:lattice}), the low energy propagation of the ECCs in a single layer graphene is characterized by the proportionality between the spatial and temporal recurrences of the wave function: $\vec \lambda  (\bar \mathbf p) = \vec v_F T  (\bar \mathbf p)$. According to (\ref{deBroglie:Planck:rel}), this is equivalent to say that the ECCs have the dispersion relation typical of a massless particle, $\bar E = v_F |\bar \mathbf p|$, being the speed of light $c$ replaced by the Fermi velocity $v_F$. Notice that the temporal recurrence tends to infinity as the momentum tends to zero: $T (\bar \mathbf p) \rightarrow \infty$ for $\bar \mathbf p \rightarrow 0$ (infinite Compton periodicity). 

This is not the case of CNs in which the quasi-particles acquire an effective mass. The effective mass is a direct consequence of the dimensional reduction of a graphene spatial dimension to form a CN. The geometry of the compactification of the effective graphene space-time is determined by a  compactification vector $\vec C_{h}$. Its modulo $C_h$ is the circumference of the curled-up dimension of the CNs. The CN effective space-time resulting from this compactification is 1D+1. 
The curled-up dimension of CNs constrains the ECCs wave function to fulfill PBCs, which in the coordinate system of the graphene lattice are  
 \beq 
 \phi(\vec r) = \phi(\vec r + \vec C_{h})~.\label{CN:PBCs}
 \eeq 
 In this section we neglect effects associated to the discrete lattice structure. This will be discussed in sec.(\ref{spectrum}). 
 
The effective mass of ECCs is given by the fact that in CNs the  compactified spatial dimension determines an effective finite Compton periodicity. That is, the ECCs wave function acquires a rest periodicity with respect to the effective motion long the CN axial direction. This can be easily seen by  considering a massless (quasi)particle moving at velocity $v_F$ along the CN curled-up dimension, so that it has a cyclic motion of tangent velocity $v_F$ along CN circumference  and  zero velocity in the axial direction\footnote{A consequence of the lattice structure is that in some CN geometries it is not possible to have a purely radial motion.}. This means that the ECC is at rest with respect to the CNs 1D+1 effective space-time, namely $p_\parallel = 0$. Therefore the quasi-particle at rest in the CN 1D+1 space-time has a residual \emph{rest} recurrence, i.e. a finite effective Compton periodicity.
  In terms of the effective proper-time coordinate $\tau $ of the quasi-particle in the CN effective 1D+1 space-time,  the PBCs (\ref{CN:PBCs}) for the cyclic motion along the CN circumference is $$\bar \psi(\tau) =  \bar \psi(\tau + T_C)\,.$$ We conclude that the the ECCs in CN have Compton time periodicity $T_C = T(p_\parallel = 0) = {C_{h}}/{v_F} $. Equivalently, this compactification implies a recurrence in the world-line parameter $s = \tau  v_F$ of the quasi-particles in the CN effective 1D+1 space-time: $\bar \psi(s) =  \bar \psi(s + C_h)$.  In turn, the mass scale of the ECCs is determined by this effective rest periodicity according to the Compton relation
\beq 
 \bar m = \frac{2 \pi \hbar}{  C_{h} v_F} = \frac{2 \pi \hbar}{ T_C v_F^2} \,.\label{Compt:period.ECC}
 \eeq
 Indeed in undulatory mechanics the mass (rest energy) corresponds, through the Planck constant, to a rest periodicity of the wave function, namely the Compton periodicity which can be expressed either as a proper-time recurrence  or a world-line recurrence. This also means that the ECCs in single layer graphene, being massless, have an infinite Compton periodicity (i.e. infinite CN compactification length).
This describes the effective mass scale in CNs in agreement with experimental observations. 
In sec.(\ref{spectrum}) we will derive the exact mass spectrum in the specific CNs configurations by considering the corresponding lattice geometry. 

\begin{figure}[htbp]
\begin{center}
\includegraphics[scale=0.8]{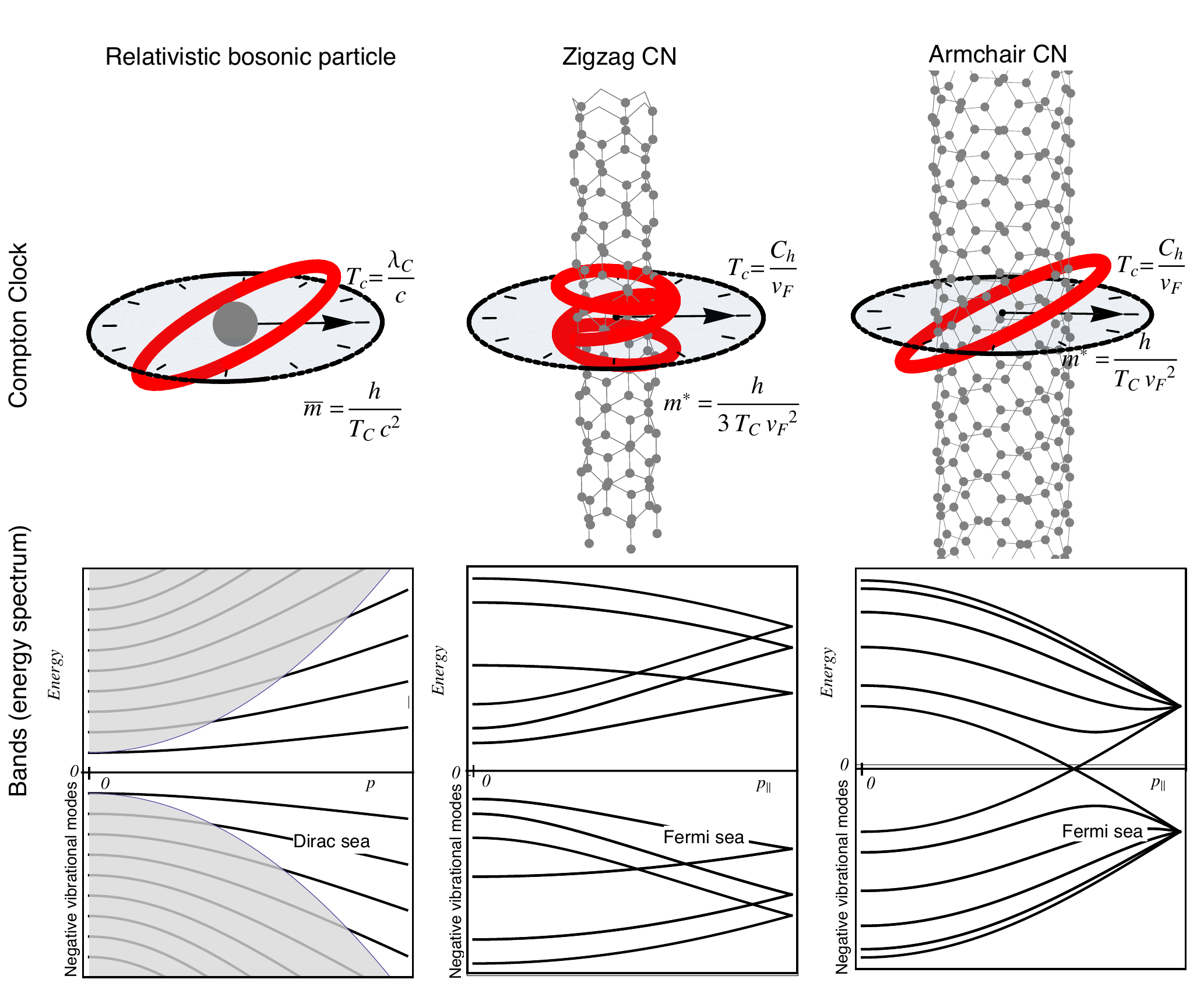}
\caption{On the upper part of the figure we report a schematic representation of the Compton clocks (black) of a free relativistic particle, and of the (quasi)particles in Zigzag (ZZ) and Armchair (AC) CNs. In CNs the characteristic period $T_C$ of the Compton clock, which for electrons is  $ \sim 10^{-21}$ s, turns out to be rescaled to $ \sim 10^{-15}$ s, i.e. to timescales accessible experimentally. Thus CNs, similarly to elementary particles, can be regarded (and potentially used) as timekeepers of resolution $ \sim 10^{-15}$ s. The red thick lines represent the wave functions of the fundamental vibrational states associated to the Compton clocks, determining the corresponding masses by means of the Compton relation. Notice that in ZZ CNs the closed orbit of the fundamental wave function has three windings as a consequence of the twist factor $ 2 \pi \alpha = \pm 2 /3 \pi$ in the rest periodicity condition, eq.(\ref{PBCs.twist:tau}). In the continuum lattice limit CNs behave as in the first column provided that the Compton wavelength $\lambda_C$ is replaced by the CN circumference $C_h$ and the speed of light $c$ by the Fermi velocity $v_F$. The bottom part of the figure reports the normal ordered quantum energy spectrum of a free particle (the gray opaque part cannot be observed for free particles with small momentum, see Sec.3) and the CNs energy bands in the ZZ and AC case, respectively. The negative modes (not reported in the figure) describe anti-particles, i.e. holes in the Fermi sea. In short, by considering the relativistic modulations of periodicity associated to the motion (e.g. the relativistic Doppler effect), the CN energy bands are the vibrational spectrum of the Compton clock.}
\label{fig.1}
\end{center}
\end{figure}

The same arguments can be extended to describe the effective ECC mass in other graphene systems. For instance, in graphene bi-layer the role of the rest recurrence is determined by the distance $d$ of the two layers. Through electromagnetic interaction, the ECCs among the two layers result strongly hybridised. ECCs at rest in the planar direction have a correlation among the layers. We denoted  by $x_\bot$ the direction perpendicular to the layers and by $\Sigma_{1,2}$ the position of the layers in this direction. In analogy with the rest periodicity in CNs, these fix the boundary conditions for the wave function (in this case  $\partial_\bot \psi(x_\bot)|_{\Sigma} = 0$,  similarly to a particle in a box or a vibrating string in which Dirichlet BCs are replaced by Neumann BCs). The Compton length, \emph{i.e.} the analogous of the compactified length of CNs, is given by the distance among the layers and the resulting spatial dynamics of massive ECCs are now bi-dimensional (2D+1) \cite{JanZaanen_2013,2011arXiv1103.1663Z,2013PhRvL.110n6803P}. This separation $d$ is of the order of $\sim 10^{-10}$m. If compared with the Compton length of the electron $\sim 10^{-12}$m we find that the effective mass of the ECCs in graphene bilayer is $\bar m \sim 10^{-2} \times m_{e}$ as confirmed by observations\footnote{{For a more accurate evaluation it is necessary to consider the reduced speed of light $\hat c$ in the motion among the two layers.}}. More complex graphene systems such as multiwalled CNs are described by the combination of these two fundamental cases so that, for example, the mass spectrum is characterised by two fundamental mass scales. Another example is provided by the superlattice of quantum stripes in which the rest periodicity with respect to direction parallel to the strips is determined by the gap L between the stripes (in this case the ECCs have a non-relativistic dispersion relation, this implies that the effective mass $m \propto 1/ L^2 $) \cite{1996SSCom.100..181P}.

\begin{table*}[htdp]
\begin{center}
\begin{small}
\begin{tabular}{||c||c|c||c||}
\hline \hline
 &  Graphene & Nanotubes & Relativistic space-time \\ \hline \hline  
Dimensions & 2D &  1D &  3D    \\ \hline
Speed of Light & $v_F$ & $v_F$ & $c$
\\ \hline 
Periodicity & $\psi(\vec r) \equiv \psi(\vec r + \vec a_1 n + \vec a_2 m)$ & $\bar \psi(\vec r) =  e^{ i 2 \pi \alpha } \bar \psi(\vec r + \vec C_{h})$  & $\phi(x^{\mu}) = e^{i 2 \pi \alpha}\phi (x^{\mu} +  \lambda^{\mu})$ \\ 
& (+  $\partial_\bot\psi(x_\bot)|_{\Sigma} = 0$ for BL) &  or $\phi(x_{\ast}^{\mu}) = e^{i 2 \pi \alpha}\phi (x_\ast^{\mu} +  \lambda_{\ast}^\mu)$
&  \\ \hline

 Compton   & $T_C = \infty$ & $T_C = \frac {C_h}  {v_F}$ & $T_C = \frac {\lambda_C} { c}$ \\
Clock &( $T_C = {d} / {2 \hat{c}} $ for BL) & & \\ \hline
 Mass Scale & $\bar m \sim 0$& $\bar m = \frac{h}{ |\vec C_{h}| v_F} = \frac{h}{ T_c v^2_F}$  & $ \bar m = \frac {h} { \lambda_C c}=  \frac {h} { T_C c^2} $  \\ 
  & ( $\bar m  \sim \frac{ h c} { d}$ for BL) &  & \\ \hline
  Dispersion &   $\bar E^{2}(\bar p_\parallel) \simeq  {\bar{p}}_\parallel^{2} v_F^{2} $ &  $\bar E^{2}(\bar p) = \frac{(2 \pi \hbar)^2}{T^{2}({\bar{p}}_\parallel)} \simeq {\bar m^2 v_F^{4} + {\bar{p}}_\parallel^{2} v_F^{2}} $ & $\bar E^{2}(\mathbf{\bar{p}}) = \frac{(2 \pi \hbar)^2}{T^{2}(\mathbf{\bar{p}})}={\bar m^2 c^{4} + \mathbf{\bar{p}}^{2} c^{2}}$ \\
  	relation & & & \\ \hline
	Phase  & & & \\
	 harmony &  $\vec{\mathbf {\bar{ p}}} \cdot (\vec a_1 n + \vec a_2 m) = 2 \pi \hbar $  & $ \vec{\bar{ p}} \cdot \vec C_{h} = {\bar p}_{\ast \mu}  \lambda_{\ast}^\mu = 2 \pi \hbar$ & ${\bar{p}}_{\mu}  \lambda^{\mu}\equiv 2 \pi \hbar$ \\ 
	 & & & \\\hline
space-time    & $0 \equiv \frac{1}{T^2(\mathbf{\bar p})} -  \sum_{i} \frac{v^2_F}{ \lambda_i^2(\mathbf{\bar p})} $ & $ \frac{1}{T^2_C} =  \frac{v_F}{\lambda_{\ast}^\mu} \frac{v_F}{\lambda_{\ast\mu}}= \frac{1}{T^2(\bar p_\parallel)} - \frac{v_F^2}{\lambda_{\parallel}^2(\bar p_{\parallel})}$  &  $ \frac{1}{T_C^{2}} = \frac{c}{\lambda^{\mu}} \frac{c}{\lambda_{\mu}} = \frac{1}{T^2(\bar \mathbf{p})} - \sum_{i} \frac{c^2}{\lambda_i^2(\bar \mathbf{p})}$ \\ 
recurrence  & & & \\ \hline
Energy  & Bloch theorem ... & Cont: $E_n(p_\parallel)  = (n + \alpha)  \sqrt{\bar m^2 v_F^4 + \bar p^2_{ \parallel} v_F^2}$  & $E_{n} (\bar {\mathbf{p}}) = \left(n + \frac 1 2\right) \sqrt{  \bar m^{2} c^{4} + \mathbf{p}^{2} c^{2}}$   \\
	spectrum & & ZZ: $  E^2_n(\bar {p}_{\parallel})  = \bar m^2 v_F^4  \frac{  N^2}{3 \pi^2} \left(1 + 2 \sin \frac{\pi n}{N}\right)^{2} - $ & (free particle)  \\
	& & $- \frac{ 2}{3} \bar {p}_{\parallel}^2 v_F^2   \cos \frac{\pi n}{N}$ & \\ \hline
Rest  & none in the low  & Cont  $m_n = (n + \alpha) \bar m$ & ${E_n(0)} = \left(n + \frac 1 2\right)  \bar m c^2 $ \\
spectrum & energy approximation & ZZ: $m_n  =  \bar m \frac{ N}{ \pi} \sin \frac{\pi n}{N}$ & (not observable for free particles)
 \\
 & & AC: $m_n  =   \bar m \frac{ N}{3 \pi} \left(1 + 2 \cos \frac{\pi n}{N}\right) $ &  \\ \hline
Spectral  & & $ \rho(E) = \mathcal R \sum_{n} \frac{|E_n(p_\parallel)|}{\sqrt{E_{n}^2(p_\parallel)- m^{2}_{n} v^{2}_{F}}} $ &  $ Im \Pi(p^2) =  \sum_n \frac{F^2_n(p^2)}{p^2 - m_n^2 c^2 } $ \\ 
Density & &  & strongly interacting (e.g. QCD) \\ \hline \hline
\end{tabular}
\end{small}
\end{center}
\caption{The table summarizes the correspondences between ECCs in graphene systems and elementary particles, emphasizing the rule in of the quantum recurrences in both cases. The graphene systems reported are single layer graphene, bilayer graphene (BL), and CNs in the continuum limit (Cont), in the Zigzag (ZZ) and in the Armchair (AC) configurations.}
\label{tab1}
\end{table*}%

A similar undulatory interpretation of the mass can be generalised to elementary particles. In this case the Fermi velocity is replaced by the speed of light $c$ which relates the mass to the rest energy according to $\bar E(0) = \bar m c^2$. On the other hand,  in undulatory mechanics, the Planck constant relates the energy $\bar E^{2}(\bar \mathbf p)$ of a particle in a generic reference frame denoted by $\bar \mathbf p$ to the particle wave function recurrence in time, $T(\bar \mathbf p)$, according to (\ref{deBroglie:Planck:rel}). 
The de Broglie relation for a rest particle implies that the Planck constant associates the particle mass $\bar m$  to a Compton periodicity $T_C=T(0)$. The generalization of (\ref{Compt:period.ECC}) for an elementary particle is
\beq
T_C=T (0) = \frac{\lambda_{C}}{c} = \frac{2 \pi \hbar }{ \bar m c^2}  = \frac{2 \pi \hbar}{ \bar E(0)}~. \label{prop:time:CN}
\eeq 

Thus every massive particle is characterised by recurrences along its world-line $s = c \tau$ whose period is the Compton wavelength of the particle $\lambda_{C} =  c T_C $.  The heavier the particles mass, the faster the characteristic period of the Compton clock. Even considering a light particles such as the electron, this Compton temporal period is extremely fast, about $T_C \sim 10^{-21} $s. This is extremely fast even if compared to the characteristic periodicity of the Cs133 clock, about  $10^{-10} $s or to the present resolution in time which is of the order of  $10^{-17}$ s (close to the attoseconds scale). On the other hand, a massless particle such as the photon has infinite proper-time periodicity, i.e. infinite Compton wavelength.  CNs have therefore the important property to effectively rescale the Compton periodicity of elementary particles to values directly observable experimentally. Since the Fermi velocity in graphene is about $v_F \sim 10^6$ m / s and the typical circumference of CNs is of the order of $C_h \sim 10^{-9}$ m, the rescaled Compton periodicity in CNs is $T_C \sim 10^{-15} $ s, which today is  experimentally accessible \cite{Lan01022013,Margolis_2014,Rusin:ZBG:CNs,Catillon:DebroglieClock}.

A direct observation of the cyclic dynamics of elementary particles is potentially relevant to understand foundational aspects of QM such as the \emph{zitterbewegung}\cite{PhysRevB.80.045416}, the emergence of the wave-particle duality and related aspects as recently proposed in \cite{Dolce:cycles,Dolce:tune,Dolce:AdSCFT,Dolce:2009ce}. Due to the Compton periodicity CNs could be used to test ``clocks directly linking time to the particle mass'', \cite{Lan01022013}, as well as to simulate and indirectly test the ``internal clock'' of elementary particles conjectured by de Broglie. 
The rescaled Compton periodicity of CNs has experimental interest to build precise timekeepers, that is to say reference clocks with time accuracy of the order of $10^{-15}$ s. 
Finally, a good modelization of the ECCs cyclic dynamics is potentially important to have an efficient control of the electronic signals in CNs ultra-fast electronic devices, which are likely to have relevant applications in next-generation hardware  \cite{citeulike:11433540}.

\section{Dispersion relation and space-time recurrence}

The dispersion relation of ECCs in CNs can be interpreted by means of space-time recurrences of the corresponding wave-function and linked to the geometry of the CNs. For the sake of simplicity, here we consider the continuum graphene lattice limit.  
Observations show that in the continuum limit the typical dispersion relation in a CN is  \cite{RevModPhys.79.677}
   \beq
 \bar E^{2}(\bar p) = {\bar m^2 v_F^{4} + {\bar{p}}_\parallel^{2} v_F^{2} + \mathcal O (\bar p^{3})}~. \label{fund:relat:disp:CNs}
 \eeq
 This approximates the dispersion relation of relativistic massive particles, provided that the speed of light $c$ is replaced by the Fermi velocity $v_F$. In the effective 1D+1 space-time of CNs, the energy-momentum is a two-vector $\bar p_{\ast \mu}$ which can be inferred  from the ECCs rest mass through Lorentz transformations: $\bar m \rightarrow \bar p_{\ast \mu}=\{\bar E/v_F, -{\bar{p}_\parallel}\}$ where  $\bar E = \gamma_\ast \bar m v_F^2$, ${\bar{p}}_\parallel = \gamma_\ast {v}_\parallel  \bar m $, $\gamma_\ast = (1- v^2_\parallel / v^2_F)^{-1/2}$ and $v_\parallel$ is the velocity of the ECCs along the CNs axis.
 
 The geometric derivations of (\ref{fund:relat:disp:CNs}) in terms of space-time recurrences is the following. The rest state is characterized by a Compton recurrence $T_C$, see (\ref{prop:time:CN}). Since $\bar p_\parallel = 0$, the ECCs at rest have an infinite spatial recurrence in the axial direction of the CN: $\lambda_\parallel = \infty$.  As the ECCs start to move along the CN from their rest frame, for instance as a consequence of a voltage gradient along the axial direction, the non vanishing  $\bar p_\parallel $ corresponds to a finite spatial recurrence along the axis, $\lambda_{\parallel}(p_{\parallel}) = 2 \pi \hbar / \bar p_{\parallel}$, namely (2) with $i = \parallel$. On the other hand, in analogy with the relativistic Doppler effect, the temporal periodicity turns out to be modulated with respect to the rest periodicity depending on its kinematical state: $T(\bar p_\parallel) = 2 \pi \hbar / E (\bar p_\parallel)$. The dispersion relation (\ref{fund:relat:disp}) implies that these space-time periodicities (in the continuum limit) are related by the constraint $\frac{1}{T_C^{2}} = \frac{1}{T(\bar p_\parallel)^2} - \frac{v_F^2}{\lambda_{\parallel}^2(\bar p_{\parallel})}$, as can be easily inferred multiplying by $(2 \pi \hbar)^2$. This means that $T(\bar p_\parallel)$ and  $\lambda_{\parallel}$ transform as the components of a contravariant space-like (tangent) two-vector $\lambda^\mu_\ast = \{T(p_\parallel) v_F, \lambda_{\parallel}\}$. That is, $\lambda^\mu$ is determined by the corresponding Lorentz transformation of the Compton recurrence: $v_F T_C = v_F \gamma_\ast T - \gamma_\ast  v_\parallel \lambda_\parallel / v_F  $. The energy-momentum is directly derivable from $\lambda_\ast^\mu$ according to the de Broglie---Planck---Einstein relation. By using this Lorentz transformation of $T_C$, see above, we find that  the relation between energy-momentum and space-time recurrence is given by the so-called phase-harmony condition 
 \beq
\bar m T_C v^2_F  =  {\bar{p}}_{\ast \mu}  \lambda_{\ast}^\mu\equiv 2 \pi \hbar~.
\eeq

We now generalise this description to the space-time recurrence of elementary particles, \cite{Dolce:cycles,Dolce:tune,Dolce:AdSCFT,Dolce:2009ce}.  The mass $\bar m$ of an elementary particle determines through Lorentz transformation the kinematical state of the particle in generic reference frame: $\bar m \rightarrow \bar p_{\mu}=\{\bar E/c, -\mathbf{\bar{p}}\}$, where $\bar E = \gamma \bar m$ and $\mathbf{\bar{p}} = \gamma \mathbf{v}  \bar m $, $\gamma$ is the Lorentz factor. On the other hand, as noted in \cite{Dolce:cycles,Dolce:tune,Dolce:AdSCFT,Dolce:2009ce},  the Lorentz transformation of the Compton recurrence (\ref{prop:time:CN}) determines the space-time recurrence of the particle wave function in the corresponding reference frame $c T_C \rightarrow c \gamma  T - \gamma \vec v \cdot \vec \lambda / c$ and they can be written as a four-vector  $\lambda^{\mu}=\{T c, \vec \lambda\}$. According to the wave-particle duality, the space-time recurrence $\lambda^{\mu}(\mathbf{\bar{p}})$ fixes through the Planck constant the local four-momentum of the particle $\bar p_{\mu}$ in the corresponding reference frame, according to (\ref{deBroglie:Planck:rel}) with $i=1,2,3$  denoting the three spatial dimensions  of ordinary relativistic space-time (3D+1).  Similarly to the case of CNs, the energy-momentum and the space-time recurrence are related by the phase-harmony condition
\beq
\bar m T_C c^2  \equiv  {\bar{p}}_{\mu}  \lambda^{\mu}\equiv 2 \pi \hbar ~.
\eeq
  This product is a relativistic invariant (the phase of a relativistic matter wave is invariant) and  $\lambda^{\mu}$ transforms as a contravariant tangent space-like four-vector \cite{Kenyon:1990fx,Dolce:cycles,Dolce:tune,Dolce:AdSCFT,Dolce:2009ce} ($\bar p_{\mu}$ is a covariant tangent four-vector).  This means that  $\lambda^{\mu}$ satisfies  the reciprocal of the relativistic dispersion relation of the four momentum: $\bar m^{2} c^2= \bar p_{\mu} \bar p^{\mu} \Leftrightarrow \frac{1}{T_C^{2}} = \frac{c}{\lambda^{\mu}} \frac{c}{\lambda_{\mu}}$ \footnote{This description is consistent as long as we want to deal with free particles. For a consistent description of interaction, local modulations of  $\lambda^{\mu}$ must be considered as described, \cite{Dolce:cycles,Dolce:tune,Dolce:AdSCFT,Dolce:2009ce}.}.  Similarly to the relativistic Doppler effect, the time recurrence of the matter wave as observed from different reference frames is modulated according to the relativistic dispersion relation  
\beq 
\bar E^{2}(\mathbf{\bar{p}}) = \frac{(2 \pi \hbar)^{2}}{T^{2}(\mathbf{\bar{p}})} = {\bar m^2 c^{4} + \mathbf{\bar{p}}^{2} c^{2}}\,. \label{fund:relat:disp}
\eeq

\section{Energy bands and quantum energy spectrum}\label{spectrum}

In this paragraph we will consider the effect of the discrete graphene lattice. Besides  the PBCs associated to the curled up dimension (\ref{CN:PBCs}) the exact energy band structure of the ECCs in CNs is determined by the lattice periodicity (\ref{graphene:lattice}). The lattice definition of the compactification vector is $\vec C_{h} = N_{1} \vec a_{1} + N_{2} \vec a_{2}$, with fixed $N_1$, $N_2$  $\in \mathbb N$. 

Due to the hexagonal  lattice of graphene some CNs configurations (those with $N_1 - N_2 = 3 N \pm 1$, $N \in \mathbb N$) are symmetric under rotations of $\pm \frac{2}{3} \pi$. In this case the PBCs (\ref{CN:PBCs}) have a twist factor
\beq 
\bar \psi(\vec r) =  e^{ i 2 \pi \alpha } \bar \psi(\vec r + \vec C_{h})~.  \label{CN:std:PBCs:twist}
 \eeq 
 As we will see below,  $\alpha = \pm \frac{1}{3} $ will corresponds to semiconducting CNs whereas the case of $\alpha = 0 $, i.e.  (\ref{CN:PBCs}), will refer to metallic CNs.
 
 The  PBCs (\ref{CN:std:PBCs:twist}) imply a quantisation of the ECCs in analogy with the ``particles in a box'' or a vibrating string. The ECCs wave function turns out to be a superpositions of all the possible harmonic modes satisfying (\ref{CN:std:PBCs:twist}), i.e. all the vibrational modes of the effective Compton clock. 
 
 In terms of the coordinate system of the graphene lattice the quantisation condition for the ECC energy-momentum coming from (\ref{CN:std:PBCs:twist}) is therefore
 \beq 
e^{ i 2 \pi \alpha}  e^{-{\frac{i}{\hbar}  \vec p_{n} \cdot \vec C_{h} }}   =  1 ~~~~~~ \Rightarrow ~~~~~~   \vec p_{n} \cdot \vec C_{h} = 2 \pi \hbar (n + \alpha)~.
 \eeq

When imposed as  constraint, such a rest periodicity $T_C$ implies a quantization of the rest energy, i.e. a mass spectrum. The wave function in the rest frame can be therefore written as superposition of mass eigenstates $\bar \psi(\tau_{\ast})  = \sum_n A_n \exp{[-i m_{n} v_F^2 \tau / \hbar]}$ whose eigenvalues are determined by the rest PBCs 
 \beq 
\bar \psi(\tau) =  e^{ i 2 \pi \alpha} \bar \psi(\tau + T_C) ~~~~~~  \Rightarrow ~~~~~~ m_n c^2 T_C = 2 \pi \hbar (n + \alpha)~. \label{P:quant:CN}
 \eeq 
Similar to a vibrating string, in the continuum lattice the resulting mass spectrum is harmonic with a shift associated to the twist factor \cite{RevModPhys.79.677,Dolce:superc}
 \beq
 m_n 	\simeq (n + \alpha) \bar m =  (n + \alpha)\frac{2 \pi \hbar}{T_C  v_F^2}\label{CNs:cont:mass}\,.
 \eeq
 
 Indeed, we find that for $\alpha = \pm \frac 1 3$ all the vibrational eigenmodes are massive. This actually reproduces semiconducting CNS. In agreement to known results we find that in this case the fundamental band has mass \beq m_\ast =  \frac{2 \pi \hbar}{3 T_C  v_F^2}\,.	\eeq  On the other hand, for $\alpha = 0$ we have a massless band, characterising the metallic behaviours of the corresponding CNs. The the fist massive band has mass $\bar m$, see (\ref{Compt:period.ECC}).
 
 The effective mass spectrum  follows by considering that the periodicity is on a lattice \footnote{E.g., when we pass from a continuum vibrating string of fundamental frequency $1 / T$ to a chain of $N$ masses and springs (string on a lattice) the harmonics spectrum modifies as $\omega_n = \frac{n}{ T} \rightarrow \frac{N}{T \pi} \sin \frac{\pi n}{N} $. }. For Zigzag ($N_1 = N$ and $N_2 = 0$) and Armchair ($N_1 = N_2 = N$) CNs  we have  \cite{deWoul:2012ed} respectively
 \beq
 m^2_n  \simeq   \frac{(2 \pi \hbar)^{2} }{T_C^2 v_F^4 } \frac{N^2}{\pi^2} \sin^2 \frac{\pi n}{N}~, \label{Armchir:spectrum}~~~~~~~~~m^2_n  	\simeq    \frac{(2 \pi \hbar)^{2} }{ T_C^2 v_F^4 } \frac{N^2}{3 \pi^2} \left(1 + 2 \cos \frac{\pi n}{N}\right)^2\,.
 \eeq
 
 A similar mass spectrum can also be observed in superlattice of quantum stripes  (notice however that the ECCs have a non-relativistic dispersion relation, so that  the mass spectrum goes like  $n^2$, that is $m_n \propto n^2/ L^2 $) \cite{1996SSCom.100..181P}.  
 
As already said, the ECC evolution can be expressed by means of the CN effective space-time coordinates $ x_{\ast}^{\mu} = \{t,  x_\parallel\}$. 
The constraint of Compton periodicity $T_C = C_h / v_F$, i.e. the rest periodicity associated to the curled-up dimensions (\ref{CN:std:PBCs:twist}), in a generic reference frame corresponds to a constraint of space-time recurrence $ \lambda_{\ast}^\mu$ for the ECCs wave function (i.e. the modulations of $\lambda_\ast^\mu$ associated to the motion along the axial direction)
 \beqa 
\bar \psi(\tau) =  e^{ i 2 \pi \alpha} \bar \psi(\tau + T_C)  
~~~~~~\rightarrow~~~~~~ \bar \psi(x^\mu_{\ast}) =  e^{- i 2 \pi \alpha}  \bar \psi( x^\mu_{\ast} +  \lambda_{\ast}^\mu)~.  \label{PBCs.twist:tau}
 \eeqa 
 By considering the corresponding relativistic dispersion relations (\ref{fund:relat:disp:CNs}) for all the CN mass eigenmodes, in the continuum limit we thus find
 \beq
 E_n^2(p_\parallel) 	\simeq  (n + \alpha)^2 \frac{(2 \pi \hbar)^{2}}{T^{2}(\bar{p}_\parallel)}  = (n + \alpha)^2  (\bar m^2 v_F^4 + \bar p^2_{ \parallel} v_F^2)\,.\label{CN:spectrum:continuum}
 \eeq
 
The lattice version leads to the effective energy bands of CNs \footnote{The CN lattice is characterised by a periodicity in the axial direction which determines Brillouin zones, i.e. periodicity in $p_\parallel$}. 
 For instance, in the Zigzag  CNs, we find \cite{deWoul:2012ed}
 \begin{equation}
 E^2_n(\bar {p}_{\parallel})  \simeq   \bar m^2 v_F^4 \frac{N^2}{3 \pi^2} \left(1 + 2 \cos \frac{\pi n}{N}\right)^{2} -  \frac{2}{3} \bar {p}^2_{\parallel} v_F^2  \cos \frac{\pi n}{N}\,.\label{ZZ:CN:spectrum}
 \end{equation}
 
  From the energy bands of ECCs also follows the density of states, which can be in general expressed by means of the mass eigenstates and  (neglecting the pseudo-spin) written as, \cite{RevModPhys.79.677}, 
 \beqa
 \rho(E) 
 = 
 \mathcal R \sum_n \int d k_\parallel d (k_\parallel - k_{\parallel n}) \frac{|E(\hbar k_\parallel)|}{ \sqrt{E^2(\hbar k_\parallel) -  m^{2} v^{2}_{F}}} 
 = \mathcal R \sum_{n} \frac{|E_n(p_\parallel)|}{\sqrt{E_{n}^2(p_\parallel)- m^{2}_{n} v^{4}_{F}}}  ~,
 \eeqa
 where $\mathcal R = \frac{\sqrt{3} a^{2}}{2 \pi \hbar  T_C v^{2}_{F} }$ and $k_{\parallel n}$ are roots of the equation $E - E_{n \parallel}(\hbar k_\parallel ) = 0 $.

 The particle physics analogous of the the CNs energy bands is the particle quantized energy spectrum.   
As proposed in recent papers \cite{Dolce:cycles,Dolce:tune,Dolce:AdSCFT,Dolce:2009ce}, 
we can imagine to impose as constraint to a relativistic particle its quantum recurrence $\lambda^{\mu}$
\beqa
\phi(x^{\mu}) = e^{-i 2 \pi \alpha}\phi (x^{\mu} +  \lambda^{\mu}) 
 ~~~~~~ \Rightarrow ~~~ ~~~ p_{n \mu} \lambda^{\mu} =  2 \pi \hbar (n + \alpha) ~.
\eeqa
For elementary free particles this is  the  analogous of the CN quantisation condition (\ref{CN:std:PBCs:twist}), where we have introduced a twist factor for completeness\footnote{In QM only the modulo of the wave function has a physical meaning: the recurrence can be defined modulo a twist factor.}. It can be regarded as the generalization to free relativistic particles of the Bohr-Sommerfeld quantization prescription of close orbits. 
For instance, the quantization of the energy spectrum associated to the time recurrence $T(\mathbf{\bar{p}})$ is a harmonic (quantized) spectrum. In the case $\alpha = 1/2$ it coincides with the quantised energy spectrum of an harmonic oscillator of period $T (\bar{ \mathbf{p}})  = 2 \pi / {\bar \omega (\bar{ \mathbf{p}})}$: that is,
$E_{n} (\bar {\mathbf{p}}) = (n + \frac 1 2 ) \frac{2 \pi \hbar}{ T (\bar{ \mathbf{p}})} =   (n + \frac 1 2) \hbar  \bar \omega (\bar{ \mathbf{p}}) $.  By following the same line used to derive the energy bands in CNs,   the relativistic modulations  (relativistic Doppler effect) of the time recurrence $T (\bar{ \mathbf{p}})$  associated to a variation of reference frame $\bar{ \mathbf{p}}$ (variation of kinematical state), see (\ref{fund:relat:disp}), leads to quantised energy spectrum
\beqa E_{n} (\bar {\mathbf{p}}) = \left(n + \frac 1 2\right) \frac{2 \pi \hbar}{ T \left(\bar{ \mathbf{p}}\right)}  =   \left(n + \frac 1 2\right) \hbar  \bar \omega \left(\bar{ \mathbf{p}}\right)  = 
 \left(n + \frac 1 2\right) \sqrt{\bar{ \mathbf{p}}^{2} c^{2} + \bar m^{2} c^{4}}~.\label{particles:spectrum}
\eeqa
 This actually is the same energy spectrum prescribed by ordinary second quantisation for bosonic particles\footnote{Indeed, second quantisation prescribes that every mode with angular frequency  $\bar \omega (\bar{ \mathbf{p}}) = \sqrt{\bar{ \mathbf{p}}^{2} c^{2} + \bar M^{2} c^{4}} / \hbar$  of a field has a quantised energy spectrum $E_{n} (\bar {\mathbf{p}}) = (n + 1/2) \hbar \bar \omega (\bar{ \mathbf{p}})$. }. The normal ordering correspond to put $\alpha = 0$.
 In this semi-classical description the vacuum energy is indeed interpreted as a twist factor in the space-time periodicity. 
 The resulting semiclassical cyclic dynamics associated to $\lambda^{\mu}$ are formally equivalent to the ordinary relativistic QM of the particles. This correspondence has been proven for both the canonical and the Feynman formulations of QM, \cite{Dolce:cycles,Dolce:tune,Dolce:AdSCFT,Dolce:2009ce}. The harmonics associated to the space-time recurrence $\lambda^\mu$ corresponds to the quantum excitations of the elementary particles\footnote{Contrarily to ECCs in which the constraint of Compton periodicity is manifestly associated to the curled-up dimension, we say that for relativistic particles the constraints of Compton periodicity, or equivalently of space-time recurrence, is \emph{intrinsic} in the space-time geometry as proposed in \cite{Dolce:cycles,Dolce:tune,Dolce:AdSCFT,Dolce:2009ce}. That is, elementary particles are \emph{intrinsically} reference clocks \cite{Broglie:1924,Penrose:cycles}.}. {CNs can be regarded as 't Hooft's Cellular Automata \cite{tHooft,hooft:2014cogwheel}, as will be described in a forthcoming paper \cite{Dolce:DICE2014}}. Indeed, the 't Hooft CA model can be regarded as a ``particle moving on a circle of period $T$''. As proven by 't Hooft its mechanics turns out to reproduce the  time dynamics of a quantum harmonic oscillator of equal period $T$, e.g. the normally ordered energy spectrum $E_n = n \hbar \omega = n  \hbar / T$. Actually, we have found that the cyclic motion of the electron along the CN circumference with $N$ carbon atoms (a ``particle on a circle'' with $N$ lattice sites) describes the CN energy bands (harmonic energy spectrum in the continuum limit) and the other quantum properties of CNs. In particular this correspondence allows us to test the fact that, as pointed out by 't Hooft (see problem of the positivity of the Hamiltonian in Cellular Automata) the negative modes turn out to describe antiparticles (holes in the Dirac sea). Indeed the negative vibrational modes associated to the effective Compton clock in CNs (not shown in Fig.1) correspond to the negative CNs energy bands, i.e. to holes in the Fermi sea.

 The evaluation of the energy spectrum (\ref{particles:spectrum}) in the rest frame leads to a quantised rest energy spectrum, i.e. a quantised mass spectrum $m_n :=: E_n (0) / c^2 = n \bar m$ (after normal ordering). It must be noticed however that free particles at rest or in the non-relativistic limit are classical free particles, so the quantum excitations cannot  be observed. Thus such a mass spectrum is typically not accessible for free non-relativistic particles whereas in the relativistic limit the higher modes are interpretable as multiparticle states  \cite{Dolce:cycles,Dolce:tune,Dolce:AdSCFT,Dolce:2009ce}. Nevertheless the analogous of such a mass spectrum can be observed in strong interacting systems of elementary particles such the quark-gluon-plasma in QCD  and they can be identified  with hadrons. In this case, analogluouly  to the ECCs in CNs or in similar condensed matter systems, the particle can be imagined as in a thermal bath so that the quantum excitations can be populated thermally. 
 The particle analogous of the density of states (provided that the parametrization is given in terms of $E$ rather than $p^2$) is the spectral density 
 \beqa
\textit{Im}[ \Pi(p^2) ]= \sum_n \int d q^2  d(q^2 - m_n^2 c^2)  \frac{F_n^2(p^2)}{p^2 - q^2}  
= \sum_n \frac{F^2_n(p^2)}{p^2 - m_n^2 c^2 }~,
\eeqa
where $F_n (p^2)$ are form factors. This is the typical form of the two point function QCD in which the hadrons are represented as quantum excitations of the same fundamental system or the harmonics of a vibrating string \cite{Son:2003et,PhysRevD.9.3471}.

  We mention that the correspondence between CNs and the quantum behaviour of elementary particles represents a novel instance of Anti de Sitter/Conformal Field Theory correspondence and Holography \cite{Witten:1998qj}. This correspondence can be easily understood by  interpreting  the curled-up dimensions of the CNs as a ``virtual extra dimensions'' responsible for the Compton clock, according to description given in \cite{Dolce:cycles,Dolce:AdSCFT,Dolce:2009ce}, and will be the subject of a dedicated paper.   
  In a forthcoming paper we will also extend the interplay between graphene and particle physics by including the analogies between gauge interactions in particles physics and the modulations of the quantum recurrence associated to deformed geometries of the graphene lattice. In this way we will investigate the generation of the (pseudo) gauge field in graphene \cite{2010PhR...496..109V,Iorio:2013ifa}. This will be possible by using  the approach proposed in \cite{Dolce:tune}. Indeed, according to the correspondences between particle and graphene physics pointed out in this paper, CNs represent a concrate physical instance of the elementary space-time cycles proposed in  \cite{Dolce:cycles,Dolce:tune,Dolce:AdSCFT,Dolce:2009ce}. 

  \section{Conclusions}
  
  In recent literature there is a growing interest in studies linking condensed matter physics to high energy physics. 
 Examples in graphene physics are given by the generation of the (pseudo) spin degrees of freedom from the triangular sub-lattice structure of the graphene \cite{PhysRevLett.106.116803} or by the origin of the (pseudo)gauge field from lattice deformations. 
In this paper we have explored new links by investigating the effective mass and electronic properties of ECCs in graphene systems in terms of the geometry constraints of the wave function. In CNs the effective mass of ECCs  is generated by the compactification of a spatial dimension of the graphene. This in fact means that the ECCs acquire a \emph{rest} periodicity which is to be identified with the Compton periodicity. Indeed, as also pointed out recently by Penrose \cite{Penrose:cycles}, elementary massive particles are reference clocks of Compton periodicity, i.e. Compton clock \cite{Dolce:cycles,Dolce:tune,Dolce:AdSCFT,Dolce:2009ce,Lan01022013}. 

In undulatory mechanics, through the Planck constant and the Compton relation, a finite rest periodicity implies a non vanishing rest mass. Consistently with the results in literature, this describes the effective mass spectrum of the carriers in CNs and other systems such as graphene bilayers, quantum stripes and multi-walled CNs. We have shown that, by considering the possible vibrational modes associated to the Compton periodicity modulations resulting for the motion,  it is possible to infer the effective space-time recurrences of the ECCs wave function, and in turn the effective energy bands structure of CNs, in agreement with \cite{RevModPhys.79.677}, see fig.{\ref{fig.1}}. Similar considerations can be generalised to elementary particles  \cite{Dolce:cycles} establishing a dictionary between graphene physics and particle physics,  summarised in Tab.(\ref{tab1}). This is useful to extend computational technics typical of particle physics to graphene physics or \emph{vice versa}, as well as to indirectly test in graphene systems aspects of advanced  high energy physics theories. 

In CNs the Compton time of electrons is effectively rescaled from $10^{-21}$ s to $10^{-15}$ s, i.e. to timescales which has recently become accessible to direct experimental observations \cite{Margolis_2014}.  In recent papers \cite{Dolce:cycles,Dolce:tune,Dolce:AdSCFT,Dolce:2009ce} it has been proposed that QM emerges as a statistical description of the ultra-fast cyclic dynamics associated to the Compton periodicity, or equivalently to the  \emph{zitterbewegung} (Schr\"odinger ``trembling motion'') for Dirac particles, see also \cite{tHooft,hooft:2014cogwheel}. Thus the direct observation of the Compton clock in CNs is of interest in testing foundational aspects of QM. Such a direct observation could also allow CNs to be used as reference ``clocks linking time to the particle mass'' in order to build timekeepers with resolution of the order of $10^{-15}$ s. 
  Graphene-based systems also have a rapidly growing interest for next-generation electronic devices in which the ultra-fast undulatory nature of ECCs plays a crucial role. The formalism proposed here turns out to be particularly powerful to describe these ultra-fast cyclic dynamics of ECCs in CNs and thus to control the properties of these devices or the related electronic manipulation of information.

\end{document}